\documentclass[12pt,a4paper,dvips]{article}
\usepackage{cite,mcite}
\usepackage{graphicx}
\graphicspath{{/l3/paper/example/figs/}}
\newlength{\capindent}
\newlength{\capwidth}
\newlength{\figwidth}
\newcommand{\icaption}[2][!*!,!]{\hspace*{\capindent}%
  \begin{minipage}{\capwidth}
    \ifthenelse{\equal{#1}{!*!,!}}%
      {\caption{#2}}%
      {\caption[#1]{#2}}
  \end{minipage}}
\def\SN{\mathrm{\stackrel{\sim}{\nu}}}

\def\a34{\cos\alpha_{34}}

\def\TO{\mathrm{\longrightarrow\ }}
\def\EE{\mathrm{\rm \;e^+e^-\;}}

\def\MM{\mathrm{\;\mu^+\mu^-\;}}
\def\TT{\mathrm{\;\tau^+\tau^-\;}}
\def\l{\mathrm{\lambda}}
\def\L{\mathrm{\Lambda}}
\def\SP{\mathrm{s^{\prime}}}
\def\SS{\mathrm{s}}

\def\SU{\stackrel{\sim}{u}}
\def\SD{\stackrel{\sim}{d}}
\def\SQ{\mathrm{\stackrel{\sim}{q}}}
\def\SN{\mathrm{\stackrel{\sim}{\nu}}}

\def\STIL{\stackrel{\sim}{S}}

\def\g{\mathrm{g}}

\def\rs{\sqrt{s}}

\def\xi{x_{i}}

\newcommand {\Be}{\begin{equation}}
\newcommand {\Ee}{\end{equation}}

\newcommand {\eqref}[1]{equation~(\ref{#1})}

\newcommand {\Figref}[1]{Figure~\ref{fig:#1}}

\renewcommand{\thefootnote}{\fnsymbol{footnote}}

\begin{document}
\begin{titlepage}
\begin{flushright} 
ETHZ-IPP PR-98-02 \\
June 23, 1998 
\end{flushright}

\vspace*{4.0cm}

\begin{center} {\Large \bf
                         FERMION-PAIR PRODUCTION ABOVE THE Z\\
\vspace*{0.15cm}
                         AND SEARCH FOR NEW PHENOMENA }

\vspace*{2.0cm}
  {\Large
  Dimitri Bourilkov\footnote{\tt e-mail: Dimitri.Bourilkov@cern.ch}
  }

\vspace*{1.0cm}
  Institute for Particle Physics (IPP), ETH Z\"urich, \\
  CH-8093 Z\"urich, Switzerland
\vspace*{4.3cm}
\end{center}

%
%
\begin{abstract}
A review of the measurements of hadron, flavour-tagged and lepton-pair
production cross-sections and lepton-pair forward-backward asymmetries
performed by the four LEP experiments ALEPH, DELPHI, L3 and OPAL
at energies between 130 and 183 GeV is given. All 183 GeV results
are preliminary. The searches by the four collaborations for new 
physics phenomena like contact interactions and compositeness,
exchange of R-parity violating sneutrinos or squarks, leptoquarks
or additional heavy gauge bosons $Z'$ are summarized. No evidence for
deviations from the Standard Model expectations is found and new or
improved limits are derived.
\end{abstract}
%
%

\vspace*{1.0cm}
\begin{center}
{\it Presented at the XXXIInd Rencontres de Moriond, Electroweak
Interactions and Unified Theories, Les Arcs, Savoie, France,
March 14-21, 1998.}
\end{center}

\end{titlepage}
\clearpage
\renewcommand{\thefootnote}{\arabic{footnote}}
\setcounter{footnote}{0}

%
%
\section*{Fermion-pair Production above the Z}
The successful running of LEP in 1995-1997 at centre--of--mass energies well
above the Z resonance allows to search for new physics beyond 
the Standard Model (SM).
Any significant deviation from the SM
predictions in the electron-positron annihilation into fermion-pairs would
herald the presence of new phenomena. The differential cross section for
fermion-pair production is described by the $\gamma$ or Z exchange amplitudes
in the s-channel, and for $\EE$ final states also in the t-channel
\Be
\rm \frac{d \sigma}{d \Omega} = |\gamma_s+Z_s+elec*(\gamma_t+Z_t) + New\;Physics\;?!|^2
\label{eq1}
\Ee
The presence of additional amplitudes will signal new physics effects and can be
observed through modifications in the measured differential cross section.
In the helicity amplitudes formalism
\Be
\rm {\mathcal A}^{ef}_{LR(RL)}(s) = ({\mathcal A}^{SM}_{LR(RL)}(s) + {\mathcal A}^{NP}_{LR(RL)}(s))\frac{t}{s}
\label{eq2}
\Ee
\Be
\rm {\mathcal A}^{ef}_{LL(RR)}(s) = ({\mathcal A}^{SM}_{LL(RR)}(s) + {\mathcal A}^{NP}_{LL(RR)}(s))\frac{u}{s}
\label{eq3}
\Ee
where ${\mathcal A}^{SM}$ (${\mathcal A}^{NP}$) are the SM (or New Phenomena) amplitudes,
L (or R) denote the helicities of the ingoing and outgoing fermion, $s$, $t$, $u$ are the
Mandelstam variables. For a generic new interaction with coupling $g$ and
typical energy (or mass) scale $\Lambda$ the differential cross section can be
decomposed as follows
\Be
\rm \frac{d \sigma}{d \Omega} = SM(s,t)+\frac{g^2}{\L^2}C_{Interference}(s,t)+\frac{g^4}{\L^4}C_{New\;Phenomena}(s,t)
\label{eq4}
\Ee
If the new phenomena are strong enough (like resonance formation), they can be observed
directly through the $C_{New\;Phenomena}$ term. Otherwise, the interference with the SM
amplitudes can lead to observable (but smaller) effects. If no deviations from the SM
are found, we can constrain $g/\L$.

%
%
\section*{Measurements of Cross Sections and Asymmetries at LEP2}
All four LEP experiments have presented results  at centre--of--mass energies
130--172 GeV. All 183 GeV results are preliminary~\cite{adlo}. The collected luminosity
ranges from 6 pb$^{-1}$ at 130 GeV to 55 pb$^{-1}$ at 183 GeV. Up to 172 GeV
the statistical errors are dominant. The 183 GeV data mark the point where
systematic effects gain in importance.

The event selection is an extension of selections on the Z peak with some important
new features:
\begin{itemize}
\vspace{-0.2cm}
 \item two distinctive event classes depending on the effective CMS energy $\sqrt{\SP}$:
  \begin{itemize}
\vspace{-0.2cm}
  \item inclusive events  - $\sqrt{\SP} / \sqrt{\SS} > 0.1$ (`return to the Z')
\vspace{-0.15cm}
  \item non-radiative events - $\sqrt{\SP} / \sqrt{\SS} > 0.85\;-\;0.90$
\end{itemize}
\vspace{-0.2cm}
 \item cross section $\sim$ 10$^{-2}$ of Z peak values
  \begin{itemize}
\vspace{-0.2cm}
   \item higher non-annihilation background (beam-wall/gas, cosmic muons)
\vspace{-0.15cm}
   \item higher 2-$\gamma$ background ($\EE\TO\EE f\bar f$)
  \end{itemize}
\vspace{-0.2cm}
 \item 4-fermion `background' ($W^+W^-$, $ZZ$, $Z\EE$, $We\nu$, ...)
\end{itemize}
\vspace{-1.0cm}
\begin{figure}[htbp]
  \begin{center}
\resizebox{0.95\textwidth}{0.400\textheight}{\includegraphics{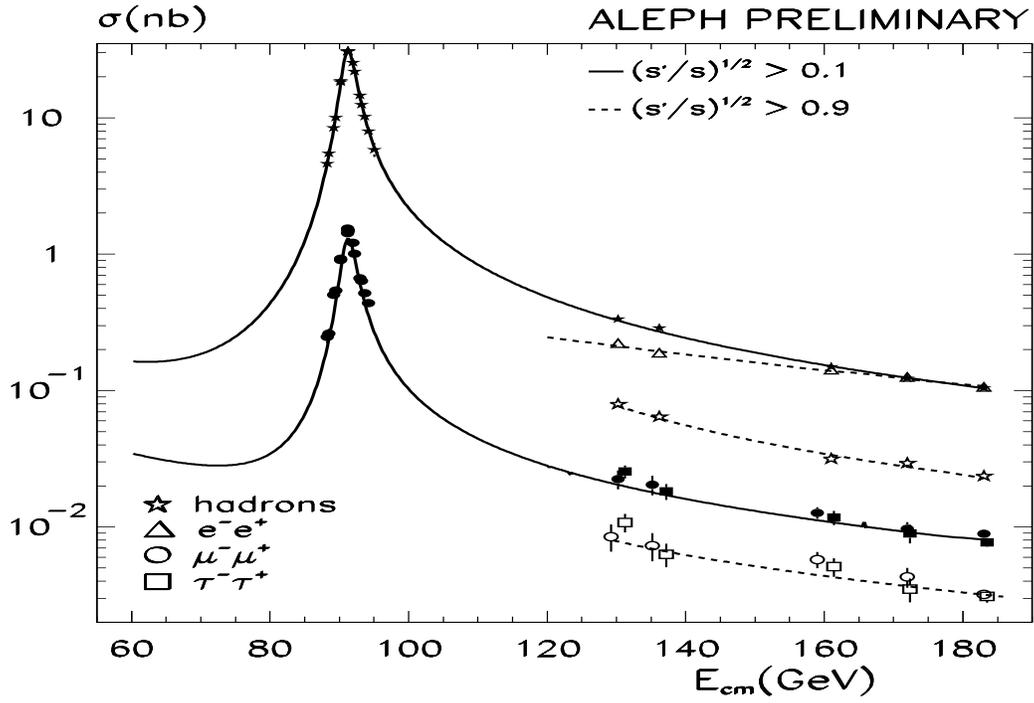}}
  \end{center}
  \vspace{-0.8cm}
  \caption{Total cross section measurements from ALEPH, including 183 GeV data.}
  \label{fig:sigmaaleph}
\end{figure}
\vspace{-2.0cm}
\begin{figure}[htbp]
  \begin{center}
\resizebox{0.90\textwidth}{0.450\textheight}{\includegraphics{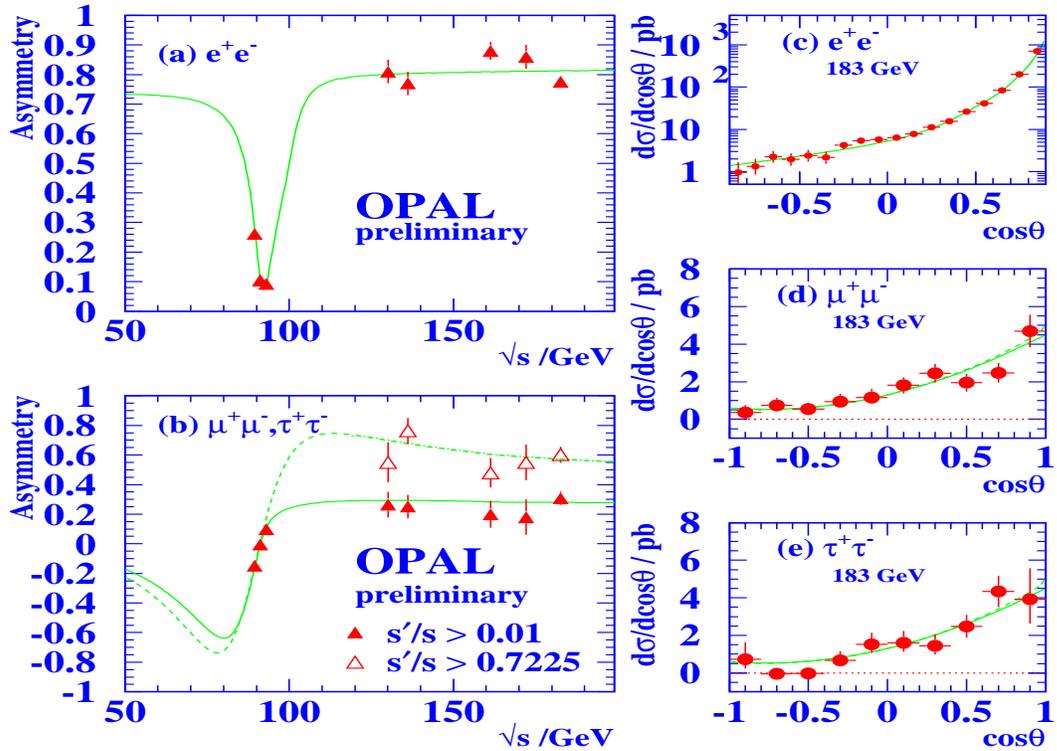}}
  \end{center}
  \vspace{-0.8cm}
  \caption{Differential cross section measurements and leptonic forward-backward
           asymmetries from OPAL, including 183 GeV data.}
  \label{fig:afbopal}
\end{figure}

Cross sections and asymmetries are measured for the inclusive and the non-radiative
events. The latter sample is the main search field for new physics. $\sqrt{\SP}$ is
calculated from the initial state radiation photon(s) if seen in the detector
or from the angles of the outgoing fermions or kinematic fits.
The definition of $\sqrt{\SP}$  is not unique due to interference between initial-
and final-state radiation. One solution (favoured by OPAL) is to correct the 
measured cross sections to `no interference'. The corrections can reach $\sim$ 1-2 \% for
$\sqrt{\SP} / \sqrt{\SS} > 0.85$ and rely on the theoretical estimate
calculated only to $\cal O (\alpha)$. A second option is an inclusive approach, where
$\sqrt{\SP}$ is defined as the effective mass of the outgoing fermion pair.

The preliminary results of all four experiments at 183 GeV show good agreement
with the SM predictions, as seen in~\Figref{sigmaaleph} and ~\Figref{afbopal}.
Applying flavour tagging for the hadron sample, ALEPH and OPAL measure the fraction
of b-quarks $R_b$ up to 183 GeV. DELPHI measures $R_b$, $R_c$ and $R_{uds}$  up to
172 GeV.

%
%
\section*{Searches for New Phenomena}
The four-fermion contact interaction (CI) offers a
general framework for a new interaction with coupling  g  and
typical energy scale  $\rm \L \gg \rs$.
By convention it is assumed that $\frac{g^2}{4\pi}=1$ and $|\eta_{ij}|\leq 1$
$(i,j\;=\;L,R)$, where $\eta_{ij}$ are the parameters defining the helicity
structure of the new interaction. The corresponding amplitudes have the
form~\cite{kroha}
\Be
\rm {\mathcal A}^{CI}_{ij}(s) = \eta_{ij}\frac{s}{\alpha}\frac{1}{\L^2}
\label{eq5}
\Ee
$\alpha$ is the electromagnetic fine structure constant.
\begin{table}[htb]
  \begin{center}
\vspace{-0.4cm}
\caption{Models of contact interaction considered. The parameters $\eta_{ij}$ define the
helicity amplitudes, $\mathcal{A}_{ij}$, which are active. Also shown are the amplitudes for
the exchange of possible new particles, where the CI gives the limiting case.}
\vspace{0.17cm}
\begin{tabular}{|c|r|r|r|r|c|c|c|}
\hline
Model     & $\eta_{LL}$ & $\eta_{RR}$ & $\eta_{LR}$ & $\eta_{RL}$ &$\ \ \ \SN\ \ \ $ &$\SQ$&Lepto-q \\
\hline
\hline
 LL       & $\rm \pm 1$ &      0      &      0      &      0      &       &$\SD$-type&$S_0(\g_L)$\\
 RR       &      0      & $\rm \pm 1$ &      0      &      0      &       &       &         \\
 LR       &      0      &      0      & $\rm \pm 1$ &      0      &$\star$&$\SU$-type&$\STIL_{1/2}$\\
 RL       &      0      &      0      &      0      & $\rm \pm 1$ &$\star$&       &         \\
 VV       & $\rm \pm 1$ & $\rm \pm 1$ & $\rm \pm 1$ & $\rm \pm 1$ &       &       &         \\
 AA       & $\rm \pm 1$ & $\rm \pm 1$ & $\rm \mp 1$ & $\rm \mp 1$ &       &       &         \\
\hline
    \end{tabular}
  \end{center}
  \label{tab:ci-models}
\end{table}
The models considered here are defined in Table~1.
In the same table the amplitudes contributing in the specific cases of
new particle exchange, considered further, are highlighted.

\begin{table}[htb]
 \renewcommand{\arraystretch}{1.05}
  \begin{center}
\vspace{-0.4cm}
\caption{Limits on $\L_+$ ($\L_-$) in TeV at 95 \% CL for $\rm l^+l^-$ and $\rm q\bar q$ final states.}
\vspace{0.17cm}
    \begin{tabular}{|c|cc|cc|cc|cc|}
\hline
$\rm l^+l^-$& \multicolumn{2}{|c|}{ALEPH}
           & \multicolumn{2}{|c|}{DELPHI}
           & \multicolumn{2}{|c|}{L3}
           & \multicolumn{2}{ c|}{OPAL}                    \\
   &~~~$\Lambda_-$~~~
         &~~~$\Lambda_+$~~~
               &~~~$\Lambda_-$~~~
                     &~~~$\Lambda_+$~~~
                           &~~~$\Lambda_-$~~~
                                  &~~~$\Lambda_+$~~~
                                         &~~~$\Lambda_-$~~~
                                                &~~~$\Lambda_+$~~~ \\
\hline  \hline
~~~~LL~~~~ & 5.4 & 6.1 & 4.4 & 3.8 & 4.1 & 5.8 & 5.3 & 5.2 \\
RR         & 5.2 & 5.9 & 4.2 & 3.7 & 4.0 & 5.6 & 5.1 & 5.0 \\
\hline  
LR         & 5.2 & 6.5 & 4.7 & 4.1 & 5.4 & 3.9 & 5.2 & 5.6 \\
\hline   
VV         & 9.4 &11.8 & 8.2 & 7.1 & 8.5 & 8.5 & 9.3 & 9.6 \\
AA         & 9.0 & 8.2 & 6.1 & 6.2 & 5.3 & 9.3 & 8.3 & 7.7 \\
\hline
\hline
$\rm q\bar q$& \multicolumn{2}{|c|}{ALEPH}
           & \multicolumn{2}{|c|}{DELPHI}
           & \multicolumn{2}{|c|}{L3}
           & \multicolumn{2}{ c|}{OPAL}                    \\
   &~~~$\Lambda_-$~~~
         &~~~$\Lambda_+$~~~
               &~~~$\Lambda_-$~~~
                     &~~~$\Lambda_+$~~~
                           &~~~$\Lambda_-$~~~
                                  &~~~$\Lambda_+$~~~
                                         &~~~$\Lambda_-$~~~
                                                &~~~$\Lambda_+$~~~ \\
\hline  \hline
~~~~LL~~~~ & 2.7 & 3.9 & tag &     & 2.5 & 3.8 & 2.8 & 4.4 \\
RR         & 3.6 & 2.9 &     &     & 3.5 & 2.8 & 3.9 & 3.0 \\
\hline  
LR         & 3.4 & 3.1 &     &     & 3.2 & 3.0 & 3.6 & 3.3 \\
RL         & 4.3 & 2.4 &     &     & 4.3 & 2.3 & 4.9 & 2.5 \\
\hline   
VV         & 5.2 & 4.0 &     &     & 5.0 & 3.8 & 5.7 & 4.1 \\
AA         & 3.7 & 5.6 &     &     & 3.5 & 5.6 & 3.8 & 6.3 \\
\hline
\end{tabular}
  \end{center}
  \label{tab:ci-fits}
\end{table}
The four collaborations express the results of their fits as 95 \% CL lower limits
on the scale $\L$ for 
$\EE$, $\MM$, $\TT$, $\rm l^+l^-$ (leptons), 
$\rm q\bar q$, $u\bar u$, $d\bar d$ (one up or down flavour is affected),
flavour-tagged $\rm q\bar q$,
$\rm f\bar f$ (all fermions combined) final states.
The limits $\L_+$ ($\L_-$) correspond to the upper (lower) sign combination in
Table~1. ALEPH, L3 and OPAL have updated their results using 183
GeV data, as shown in Table~2. Only the combined $\rm l^+l^-$
and $\rm q\bar q$ results are given, when available. 
%

The fermion-pair cross section can be sensitive to exchange of single
supersymmetric particles~\cite{kalinowski,*zerwas}
if R-parity is violated ($\rm \not\! R_{p}$).
The most general superpotential even for a minimal supersymmetric model
contains interactions violating the lepton number L and coupling sparticles
to lepton-lepton or lepton-quark pairs
\Be
\rm W_{\not R} = \lambda_{ijk}L^i_L L^j_L \bar{E}^k_R + \lambda^{\prime}_{ijk}L^i_L Q^j_L \bar{D}^k_R + ...
\label{eq6}
\Ee
where (L,Q) are left-handed doublets of leptons, quarks,
(E,D) are right-handed singlets of charged leptons, down-type quarks and
{\sl ijk} are generation indices.
These terms introduce 9 $\l$ and 27 $\l^{'}$ independent Yukawa couplings
(as $\lambda_{ijk} \not = 0$ only for $i \not = j$).
Lower energy measurements place (in some cases severe) constrains on the
possible values of $\l$ and $\l^{'}$ as a function of the sparticle mass.
We will assume that one (or two) $\l$ are much stronger than the others
in the searches that follow. For heavy sparticle masses the LEP2 data
reaches higher sensitivity in some cases.

\begin{figure}[htbp]
  \begin{center}
\makebox[8.4cm][c] {\resizebox{0.49\textwidth}{0.400\textheight}{\includegraphics{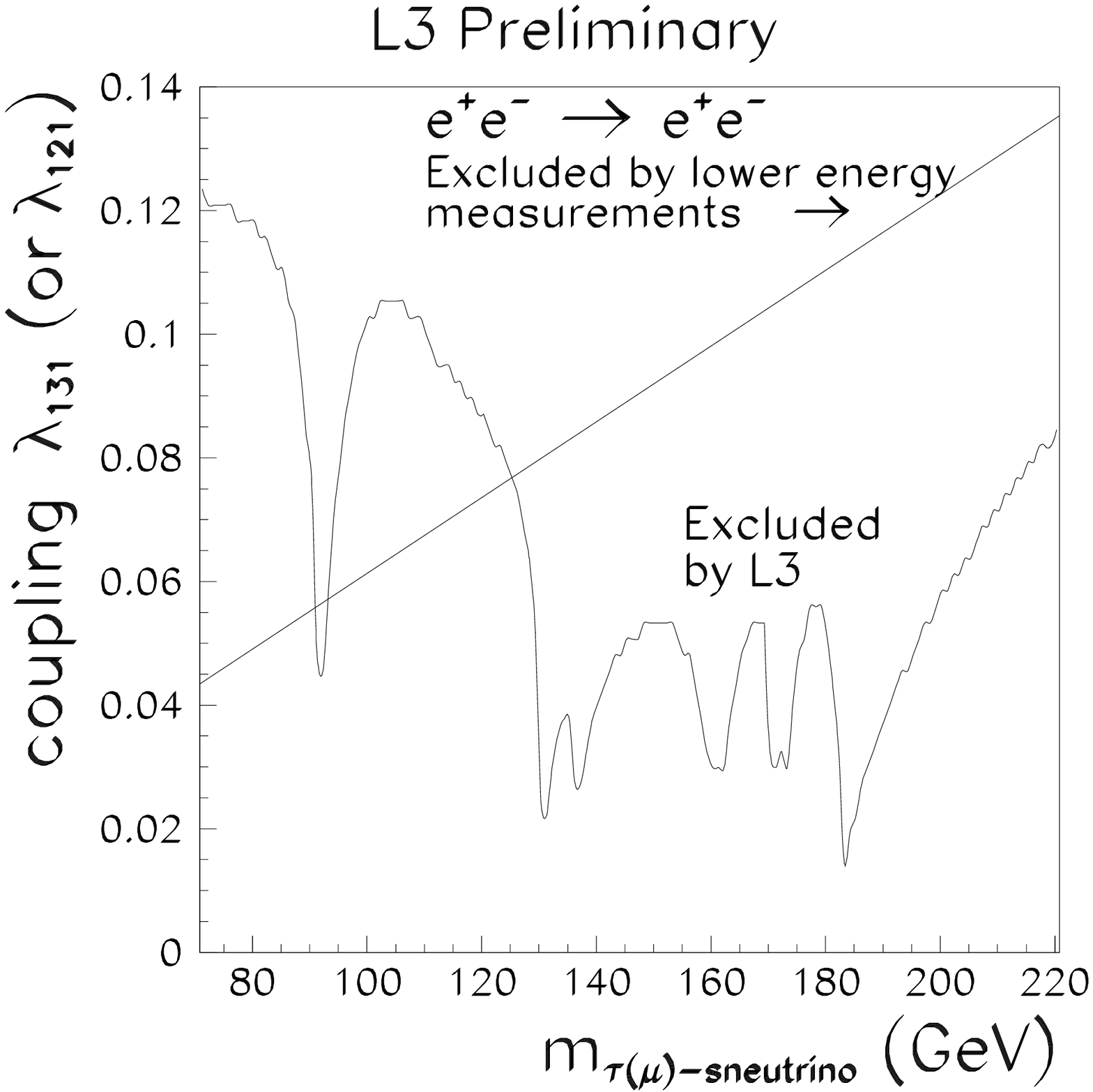}}}
\makebox[8.4cm][c] {\resizebox{0.49\textwidth}{0.400\textheight}{\includegraphics{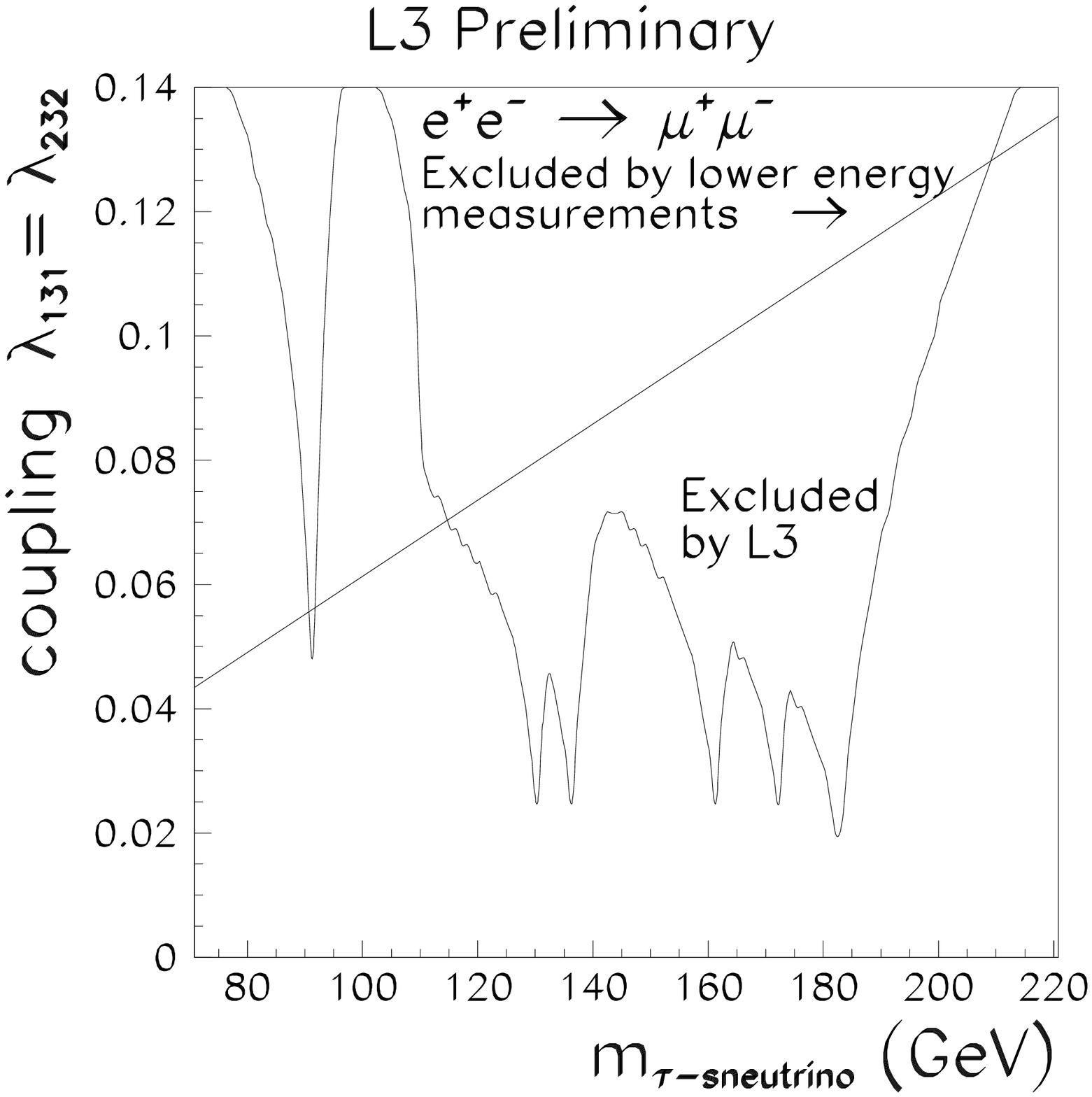}}}
  \end{center}
  \vspace{-0.8cm}
  \caption{Limits on the couplings $\l_{121}$, $\l_{131}$, $\l_{232}$ as function
           of the sneutrino mass at 95 \% CL.}
  \label{fig:snl3}
\end{figure}
The most exciting possibility is resonance formation of single sneutrinos
$\SN$ if the sneutrino mass is close to the centre--of--mass energy. Then
large effects can occur. Also t- or u-channel exchange of single $\SN$ or
squark $\SQ$ is possible in $\rm \not\! R_{p}$ theories. L3, OPAL and DELPHI
have searched for the most dramatic s-channel effects involving $\SN$.
In~\Figref{snl3} the L3 results including 183 GeV data are shown
for $\EE$ and $\MM$ final states.
The t- or u-channel $\SQ$ exchange leads to similar effects as the
exchange of leptoquarks, which appear in many unification theories.
The CDF and D{\O} collaborations exclude scalar leptoquarks with mass
below 225 GeV and branching \mbox{$\rm \beta(LQ\rightarrow l^{\pm}q) = 1$}.
At LEP2 higher masses can be probed if the coupling is sizeable.
From the many possible leptoquark states two are of special interest, as
they give limits also for squarks:
\begin{itemize}
\vspace{-0.2cm}
 \item $\rm S_0(L)$ - limit on $\l'_{1jk}$, ($j=1,2$) - 
        final state $u-type$, ($k=1,2,3$) - exchanged $\tilde d_R-type$
\vspace{-0.2cm}
 \item $\rm \STIL_{1/2}$ - limit on $\l'_{1jk}$, ($j=1,2,3$) -
        exchanged $\tilde u_L-type$, ($k=1,2,3$) - final state $d-type$
\end{itemize}
L3 and OPAL have updated their limits using 183 GeV data. For an exchanged
mass of 200 GeV typical 95 \% CL upper limits for the couplings $g$ or $\l'$
are  0.25 for $\rm S_0(L)$ and 0.65 for $\rm \STIL_{1/2}$.

A complimentary search for an additional heavy gauge boson $\rm Z'$ is
performed by the L3 and DELPHI collaborations. For specific models of the
$\rm Z'$ couplings exclusion limits in the plane of the $\rm Z'$ mass and
the mixing angle $\theta_M$ with the SM Z boson are derived.
For most of the models typical values are $\rm |\theta_M| < 0.003$
and $\rm m_{Z'} > 300 \;GeV$.

%
%
\section*{Outlook}
The production of fermion-pairs above the Z resonance is a precise testing ground,
where the Standard Model can be probed to the largest momentum transfers,
and a lively hunting field for new particles in the several hundred GeV mass
range, which couple to lepton-lepton or lepton-quark pairs. In the four-fermion
contact interaction framework the measurements are sensitive to scales approaching 
10 TeV. 

In summary:
\begin{itemize}
\vspace{-0.2cm}
  \item the Standard Model describes all fermion-pair production measurements 
        up to 183 GeV
\vspace{-0.2cm}
  \item no hint for new phenomena so far
\vspace{-0.2cm}
  \item  new or improved limits obtained: 
        contact interactions, $\SN$/$\SQ$, leptoquarks, $Z'$
\vspace{-0.2cm}
  \item the LEP2 discovery potential is large; the LEP community is looking forward to
   more data at higher energies
\end{itemize}

%

\end{document}